\documentclass[aps,prc,reprint,superscriptaddress,floatfix,showpacs]{revtex4-1}

\usepackage{amsfonts,amssymb,amsmath}
\usepackage{bm,graphicx}
\begin{document}

\preprint{}

\hyphenation{TRIUMF EDSKT DKT HENM}

\title{\boldmath Unstable Spin-Ice Order in the Stuffed Metallic Pyrochlore Pr$_{2+x}$Ir$_{2-x}$O$_{7-\delta}$}

\author{D.~E.~MacLaughlin}\email[Email: ]{macl@physics.ucr.edu.} 
\affiliation{Department of Physics \& Astronomy, University of California, Riverside, California 92521, U.S.A.}
\affiliation{Institute for Solid State Physics, University of Tokyo, Kashiwanoha 5-1-5, Kashiwa, Chiba 277-8581, Japan.}
\author{O.~O.~Bernal} 
\affiliation{Department of Physics \& Astronomy, California State University, Los Angeles, California 90032, U.S.A.}
\author{Lei~Shu}
\affiliation{Department of Physics \& Astronomy, University of California, Riverside, California 92521, U.S.A.}
\affiliation{State Key Laboratory of Surface Physics, Department of Physics, Fudan University, Shanghai 200433, China}
\affiliation{Collaborative Innovation Center of Advanced Microstructures, Fudan University, Shanghai 200433, China}
\author{Jun~Ishikawa}
\author{Yosuke~Matsumoto}
\affiliation{Institute for Solid State Physics, University of Tokyo, Kashiwanoha 5-1-5, Kashiwa, Chiba 277-8581, Japan.}
\author{J.-J.~Wen}
\altaffiliation[Present address: ]{Department of Applied Physics, Stanford University, Stanford, CA 94305, USA.}
\affiliation{Institute for Quantum Matter and Department of Physics and Astronomy, The Johns Hopkins University, Baltimore, MD 21218, USA}
\author{M.~Mourigal}
\altaffiliation[Present address: ]{School of Physics, Georgia Institute of Technology, Atlanta, GA 30332, USA.}
\affiliation{Institute for Quantum Matter and Department of Physics and Astronomy, The Johns Hopkins University, Baltimore, MD 21218, USA}
\author{C.~Stock}
\altaffiliation[Present address: ]{School of Physics and Astronomy, University of Edinburgh, Edinburgh EH9 3FD, Scotland.}
\affiliation{Institute for Quantum Matter and Department of Physics and Astronomy, The Johns Hopkins University, Baltimore, MD 21218, USA}	
\affiliation{NIST Center for Neutron Research, National Institute of Standards and Technology, Gaithersburg, MD 20899, USA}
\author{G.~Ehlers}
\affiliation{Quantum Condensed Matter Division, Neutron Sciences Directorate, Oak Ridge National Laboratory, Oak Ridge, TN 37831, USA}
\author{C.~L.~Broholm}
\affiliation{Institute for Quantum Matter and Department of Physics and Astronomy, The Johns Hopkins University, Baltimore, MD 21218, USA} 
\affiliation{NIST Center for Neutron Research, National Institute of Standards and Technology, Gaithersburg, MD 20899, USA}  
\affiliation{Quantum Condensed Matter Division, Neutron Sciences Directorate, Oak Ridge National Laboratory, Oak Ridge, TN 37831, USA}
\affiliation{Department of Materials Science and Engineering, The Johns Hopkins University, Baltimore, MD 21218, USA} 
\author{Yo~Machida}
\altaffiliation[Present address: ]{Department of Physics, Tokyo Institute of Technology, Meguro 152-8551, Japan.}
\author{Kenta~Kimura}
\affiliation{Institute for Solid State Physics, University of Tokyo, Kashiwanoha 5-1-5, Kashiwa, Chiba 277-8581, Japan.}
\author{Satoru~Nakatsuji}\email[Email: ]{satoru@issp.u-tokyo.ac.jp.}
\affiliation{Institute for Solid State Physics, University of Tokyo, Kashiwanoha 5-1-5, Kashiwa, Chiba 277-8581, Japan.}
\affiliation{PRESTO, Japan Science and Technology Agency (JST), 4-1-8 Honcho Kawaguchi, Saitama 332-0012, Japan}
\author{Yasuyuki~Shimura} 
\author{Toshiro~Sakakibara} 
\affiliation{Institute for Solid State Physics, University of Tokyo, Kashiwanoha 5-1-5, Kashiwa, Chiba 277-8581, Japan.}

\date{\today}

\begin{abstract} 
Specific heat, elastic neutron scattering, and muon spin rotation ($\mu$SR) experiments have been carried out on a well-characterized sample of ``stuffed'' (Pr-rich) Pr$_{2+x}$Ir$_{2-x}$O$_{7-\delta}$. Elastic neutron scattering shows the onset of long-range spin-ice ``2-in/2-out'' magnetic order at $T_M = 0.93$~K, with an ordered moment of 1.7(1)$\mu_\mathrm{B}$/Pr ion at low temperatures. Approximate lower bounds on the correlation length and correlation time in the ordered state are 170~\AA\ and 0.7~ns, respectively. $\mu$SR experiments yield an upper bound~2.6(7)~mT on the local field~$B_\mathrm{loc}^{4f}$ at the muon site, which is nearly two orders of magnitude smaller than the expected dipolar field for long-range spin-ice ordering of 1.7$\mu_B$ moments (120--270~mT, depending on muon site). This shortfall is due in part to splitting of the non-Kramers crystal-field ground-state doublets of near-neighbor Pr$^{3+}$ ions by the $\mu^+$-induced lattice distortion. For this to be the only effect, however, $\sim$160 Pr moments out to a distance of $\sim$14~\AA\ must be suppressed. An alternative scenario, which is consistent with the observed reduced nuclear hyperfine Schottky anomaly in the specific heat, invokes slow correlated Pr-moment fluctuations in the ordered state that average $B_\mathrm{loc}^{4f}$ on the $\mu$SR time scale (${\sim}10^{-7}$~s), but are static on the time scale of the elastic neutron scattering experiments (${\sim}10^{-9}$~s). In this picture the dynamic muon relaxation suggests a Pr$^{3+}$ $4f$ correlation time of a few nanoseconds, which should be observable in a neutron spin echo experiment.

\end{abstract}


\maketitle

\section{\label{sec:intro}Introduction}

Geometrically frustrated systems, including pyrochlore oxides, have been extensively studied because of possible novel phenomena arising from suppression of conventional order. The series of rare-earth iridate pyrochlores~\textit{R}$_2$Ir$_2$O$_7$~\cite{MoBe61} shows a nonmetal-metal transition with increasing rare-earth ionic radius~\cite{YaMa01}. The compounds with \textit{R} = Yb, Ho, Dy, Tb, Gd, and Y are nonmetallic, and those with \textit{R} = Eu, Sm, and Nd have metal-insulator transitions to antiferromagnetic ground states~\cite{MWNY07}. Only Pr$_2$Ir$_2$O$_7$, with the largest rare-earth ionic radius among the known pyrochlore iridates, remains metallic down to low temperatures (at least 50~mK)\@. Novel ground states such as spin ices and spin liquids have been proposed in the insulating pyrochlore magnets~\cite{[{For a review, see }]GGG10}. 

In the metallic pyrochlore~Pr$_2$Ir$_2$O$_7$ the Pr$^{3+}\ (J=4)$ crystalline electric field (CEF) ground state is a non-Kramers doublet that is well isolated from higher CEF levels and consists of almost pure $|{\pm}4\rangle$ states with a magnetic moment of ${\sim}3.0\mu_{\mathrm{B}}$~\cite{MNTT05}. The anisotropic field dependence of the magnetization indicates the Pr$^{3+}$ $4f$ moments have Ising-like anisotropy along the $\langle111\rangle$ easy directions. The dc susceptibility above 100~K yields an antiferromagnetic Weiss temperature~$T^{\ast} = -20$~K that has been attributed to RKKY interactions between Pr$^{3+}$ $4f$ moments~\cite{MNTT05}. 

In stoichiometric samples of Pr$_2$Ir$_2$O$_7$ neither the specific heat nor the dc magnetization exhibit any sign of long-range ordering down to a field-cooled/zero-field-cooled bifurcation temperature $T_f = 0.12$~K, where the moments partially freeze~\cite{NMMT06}. The large ratio $|T^{\ast}|/T_f = 170$ clearly indicates strongly frustrated magnetism. Between $T_f$ and ${\sim}2$~K $\chi(T)$ shows an anomalous $-\ln T$ dependence. This divergence excludes the possibility that the non-Kramers ground doublets are uniformly split into nonmagnetic singlets, and leaves open the possibility that the $4f$ moments are strongly fluctuating even for $T \ll |T^{\ast}|$, perhaps with liquid-like short-range order~\cite{NMMT06}. 

Hall-effect measurements in Pr$_2$Ir$_2$O$_7$ reveal highly unusual behavior~\cite{MNMT07,MNOT10}. The Hall resistivity, like the susceptibility, exhibits a $-\ln T$ temperature dependence, and the Hall conductivity varies strongly and non-monotonically with applied magnetic field. This behavior has been attributed to the spin chirality of Pr$^{3+}$ tetrahedral moment configurations, together with spin-dependent scattering of electrons in Ir-derived conduction bands~\cite{MNMT07}. Recently the Hall effect has been observed in Pr$_2$Ir$_2$O$_7$ in zero field and in the absence of any uniform magnetization~\cite{MNOT10}. A state of broken time reversal symmetry without conventional magnetic order is signaled by this very unusual behavior. It has been taken as evidence that Pr$_2$Ir$_2$O$_7$ is a chiral spin liquid, where the primary order parameter is chirality that is not induced by magnetic order or an applied field. 

Thermodynamic and transport properties of ``stuffed'' (Pr-rich) Pr$_{2+x}$Ir$_{2-x}$O$_{7-\delta}$~\cite{MNOM10,KON11,*KON12} reveal a well-defined phase transition at $T_M \approx 0.8$~K at ambient pressure and zero magnetic field. The transition is not found in stoichiometric samples, and is suppressed by applied field, pressure, and annealing in an oxygen atmosphere. This behavior is reminiscent of order from disorder. It is nevertheless somewhat surprising, since structural disorder would be expected to lift the non-Kramers degeneracy of the Pr$^{3+}$ crystalline electric field (CEF) ground states and thus suppress their magnetic moments.

Muon spin relaxation ($\mu$SR) experiments~\cite{[{For a review of the $\mu$SR technique and its applications see }] [{ and references therein.}] YaDdR11} have been carried out in $\mathrm{Pr_2Ir_2O_7}$~\cite{MOMN09,MNOM10} to probe local magnetic fields and their fluctuations. In zero field the muon spin relaxation function exhibits a conventional two-component Kubo-Toyabe (K-T) form~\cite{KuTo67,HUIN79} with a quasistatic~\footnote{A component~$\protect\langle\mathbf{B}_\mathrm{loc} \protect\rangle$ of the muon local field is quasistatic if it fluctuates slowly compared to the muon Larmor frequency in $\protect\langle\mathbf{B}_\mathrm{loc} \protect\rangle$. We include the static limit in our use of this term.} muon relaxation rate~$\Delta$. This behavior is often associated with nuclear dipolar fields. In $\mathrm{Pr_2Ir_2O_7}$, however, $\Delta$ is enhanced at low temperatures by one to two orders of magnitude over values expected from (predominantly $^{141}$Pr) nuclei. As discussed in Sec.~\ref{sec:quasistatic}, this enhancement is attributed to muon-induced splitting of the non-Kramers crystal-field ground state doublet of near-neighbor Pr$^{3+}$ ions~\cite{MOMN09,MNOM10}, which gives rise to hyperfine-enhanced $^{141}$Pr nuclear magnetization (HENM)~\cite{Blea73,*Blea90}. HENM has recently been observed in other Pr-based pyrochlores~\cite{FLML15}. As expected from this scenario, the temperature dependence of $\Delta$ tracks that of the local susceptibility~\cite{MNOM10}.

The present paper reports results of specific heat, elastic neutron scattering, and $\mu$SR experiments on a single sample of stuffed Pr$_{2+x}$Ir$_{2-x}$O$_{7-\delta}$. Neutron Bragg diffraction shows the onset of long-range spin-ice ``2-in/2-out'' magnetic order~\cite{MdHG01,Ging11} at $T_M$, with an ordered moment of 1.7$\mu_\mathrm{B}$ (Sec.~\ref{sec:neu}). $\mu$SR spectra taken below a magnetic transition temperature would be expected to exhibit muon spin precession in a static local field~$\mathbf{B}_\mathrm{loc}$ at the muon site~\footnote{Relaxation due to dephasing in a distribution of fields is expected if the magnetic structure is incommensurate or disordered.}. Data from the present sample, which are in agreement with earlier $\mu$SR experiments~\cite{MOMN09,MNOM10}, yield a much smaller upper bound ($\sim 2.6$~mT) on $B_\mathrm{loc}$ than would be expected from long-range spin-ice order of 1.7$\mu_\mathrm{B}$ moments (120-270~mT, depending on the muon site). 

This shortfall has also been attributed to the muon-induced Pr$^{3+}$ near-neighbor ground-state doublet splitting, which suppresses neighboring Pr moments (the ``suppressed-moment'' scenario)~\cite{MNOM10}. If this were the only effect, however, the ``suppression volume'' must contain $\sim$160 suppressed Pr moments to account for the low value of $B_\mathrm{loc}$. This number seems large for lattice distortion due to a point defect, particularly in a metal where the muon charge is screened. Furthermore, impurities in systems near a magnetic instability tend to enhance local magnetism rather than suppressing it~\cite{PRYP98,TSH99,TaHu02,XABB00}.

An alternative ``fluctuating-moment'' scenario, which is consistent with the observed reduced nuclear Schottky anomaly~\cite{BBBH02} in the specific heat of this sample (Sec.~\ref{sec:Bonville}), invokes spatially-correlated slow Pr-moment fluctuations in the ordered state. These average the local field at muon sites over the muon time scale (${\sim}10^{-7}$~s), but are quasistatic on the time scale of the elastic neutron scattering experiments (${\sim}10^{-9}$~s). Although such fluctuations seem difficult to reconcile with the long ordered-moment correlation length obtained from elastic neutron scattering Bragg peak widths ($\gtrsim 170$~\AA), similar behavior has been observed previously in the rare-earth pyrochlore stannates~Gd$_2$Sn$_2$O$_7$ and Tb$_2$Sn$_2$O$_7$~\cite{BBBH02,MAR-CB05,BMS08} and other pyrochlores~\cite{Bonv10}.

\section{\label{sec:expt}Experiment}

\subsection{\label{sec:sample}Sample synthesis and characterization}

The polycrystalline sample of Pr$_{2+x}$Ir$_{2-x}$O$_{7-\delta}$ was prepared as described previously~\cite{MMNM07,KON11,KON12}. Appropriate amounts of Pr$_6$O$_{11}$ (99.9\%), and IrO$_2$ ($>$99.9\%) were well mixed and pressed into a pellet. The pellet was wrapped in a Pt foil, placed in a silica tube, sealed under vacuum, and then fired at 1423 K for about 5 days with several intermediate grindings. Powder x-ray diffraction confirmed the single pyrochlore phase ($Fd\overline{3}m$) of the sample. Diffraction peaks were visible from small amounts of impurities ($\lesssim$10~wt.\%) identified as Pr$_3$IrO$_7$, IrO$_2$, Ir, and SiO$_2$; the latter is probably from the silica tube. The fraction of Pr$_{2+x}$Ir$_{2-x}$O$_{7-\delta}$ was estimated at 89.5(1)~wt.\%.

Scanning electron microscopy coupled with energy dispersive x-ray analysis was used to determine the composition, yielding $x = 0.4(3)$. Despite the large error, due to the polycrystalline form of the samples and the impurity phases, these results are consistent with excess Pr. Furthermore, the lattice constants of all polycrystalline samples investigated are larger than those of single crystals, which appear to grow with integer stoichiometry. This increase is also consistent with excess Pr, because the ionic radius of Pr$^{3+}$ is greater than that of Ir$^{4+}$. Thus the stoichiometry of polycrystalline samples appears to be Pr$_{2.4}$Ir$_{1.6}$O$_{7-\delta}$.

\subsection{\label{sec:spht}Specific heat}

\begin{figure}[ht]
\includegraphics[clip=,width=3.25in]{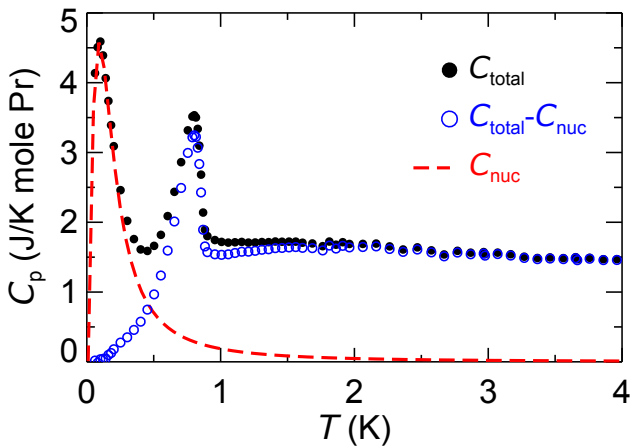}
\caption{\label{fig:sp-ht} (color online) Temperature dependence of the specific heat of Pr$_{2+x}$Ir$_{2-x}$O$_{7-\delta}$ in zero field. Filled circles: experimental total specific heat. Dashed curve: calculated specific heat due to a reduced nuclear Schottky anomaly (see text). Open circles: specific heat after subtracting the nuclear Schottky anomaly from the total specific heat.}
\end{figure}

\paragraph{Experiment.}For the specific heat measurement polycrystalline Pr$_{2+x}$Ir$_{2-x}$O$_{7-\delta}$ and silver powder for thermal contact were thoroughly mixed with approximately 1:1 mass ratio and pressed into a solid pellet. The heat capacity of this sample was measured over the temperature range~50~mK--4~K by the adiabatic relaxation method, using a Quantum Design Physical Property Measurement System with the Dilution Refrigerator option. The heat capacity of Pr$_{2+x}$Ir$_{2-x}$O$_{7-\delta}$ was then obtained by subtracting the known silver contribution~\cite{Mart73}. The temperature dependence of the specific heat~$C_p$ of Pr$_{2+x}$Ir$_{2-x}$O$_{7-\delta}$ in zero field is shown in Fig.~\ref{fig:sp-ht}. 

\paragraph{Analysis.}The resulting $C_p(T)$ shows a sharp peak at $T_M \approx 0.8$~K, with a characteristic $\lambda$ shape indicating a bulk phase transition. We associate this transition with the ordering of the Pr$^{3+}$ moments. At lower temperatures another peak in the specific heat is observed, which is attributed to a $^{141}$Pr nuclear Schottky anomaly associated with the hyperfine field~$B_\mathrm{hf}$ due to ordered Pr$^{3+}$ ionic moments. Assuming $B_\mathrm{hf}$ is static, the peak position~$T_S = 0.1$~K and amplitude~$C_p(T_S) = 4.6$~J/K mole Pr of the Schottky anomaly determine, respectively, an ordered Pr$^{3+}$ moment $\mu_S = 1.7(1)\mu_\mathrm{B}$/Pr ion on a fraction~$f = 0.65(1)$ of the Pr sites. This moment value is the same as found from elastic neutron scattering (Sec.~\ref{sec:neu}): $\mu_S = \mu_\mathrm{neu} = \mu_\mathrm{Pr}$. Such agreement is difficult to understand if a fraction~$1-f$ of the Pr$^{3+}$ ions are not ordered, since then the neutron scattering intensity would be correspondingly decreased. The reduction of the Schottky anomaly amplitude but not the ordered moment is discussed further in Sec.~\ref{sec:Bonville}.

\subsection{\label{sec:neu}Elastic neutron scattering}

\setcounter{paragraph}{0}

\paragraph{Experiment.}Powder elastic and inelastic neutron scattering data were taken from the same Pr$_{2+x}$Ir$_{2-x}$O$_{7-\delta}$ powder sample on the SPINS Triple Axis Spectrometer at the NIST Center for Neutron Research (NCNR) and on the Cold Neutron Chopper Spectrometer (CNCS) at Oak Ridge National Laboratory (ORNL)~\cite{SNAD-S14}. In both experiments the powder sample was enclosed in an aluminum can and cooled in $^3$He cryostats to base temperatures of $\sim$0.3~K (ORNL) and $\sim$0.5~K (NCNR)\@. The can was sealed under $^4$He atmosphere at room temperature to provide thermal contact for the powder. The can had an annular insert in order to minimize the effects of the strong neutron absorption in Ir. On SPINS, measurements were taken with a neutron wavelength of $\lambda = 4.04$~\AA\ ($E_i = E_f = 5$~meV), with a cooled Be filter in the incoming beam and $80'$\! collimation before and after the sample. On CNCS, measurements were taken with two neutron wavelengths, $\lambda = 7.26$~\AA\ ($E_i = 1.55$~meV) and $\lambda = 9.04$~\AA\ ($E_i = 1.00$~meV). The corresponding full-width-half-maximum (FWHM) energy resolutions at the elastic line were $\gamma=$~0.024(2)~meV and $\gamma=$~0.017(1)~meV for $\lambda = 7.26$~\AA\ with $\lambda = 9.04$~\AA, respectively. The data were normalized to absolute units using the intensity of the $(111)$ nuclear Bragg peak.

\begin{figure}[ht]
\includegraphics[width=3.25in]{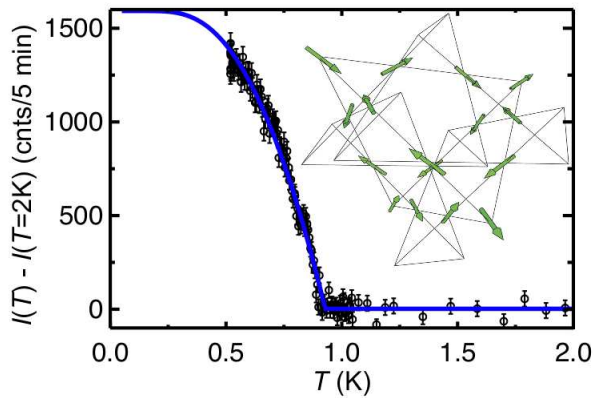}
\caption{(color online) Temperature dependence of elastic neutron scattering intensity of Pr$_{2+x}$Ir$_{2-x}$O$_{7-\delta}$ at the position of the $\mathbf{q}_m = (100)$ reflection. The intensity measured at $T = 2$~K was subtracted as a background. Curve: Ising mean-field theory fit to the data, which yields a transition temperature of $T_M = 0.93(1)$~K\@. Inset: sketch of the 2-in/2-out magnetic structure.}
\label{fig_spins}
\end{figure}

\paragraph{Analysis.}The momentum dependence of the elastic intensity was measured on SPINS over the temperature range~0.5--2~K\@. The lattice constant of the cubic space group was refined at 2~K to obtain $a = 10.672(1)$~\AA\@. Extra Bragg peaks were observed below $\sim$0.9~K, and are attributed to the ordering of the Pr$^{3+}$ moments. Their positions can be indexed using a magnetic propagation wave vector $\mathbf{q}_m = (100)$ in reciprocal lattice units of the $Fd\overline{3}m$ space group. The temperature dependence of the first peak, for which $Q = |\mathbf{q}_m|$, is shown in Fig.~\ref{fig_spins}. 
The order parameter increases continuously below $T_M$, suggesting a second-order phase transition. The data are, however, also consistent with a heterogeneous distribution of first-order phase transitions. An Ising mean-field theory provides an acceptable fit to the data (solid curve in Fig.~\ref{fig_spins}) with an ordering temperature $T_M = 0.93(1)$~K. 

Refinement of the magnetic structure using the propagation vector $\mathbf{q}_m$ was carried out on the high-temperature-subtracted $T = 0.5$~K data collected on SPINS\@. Assuming an Ising anisotropy in the $[111]$ direction for Pr$^{3+}$ moments, as is well established for Pr$_2$Ir$_2$O$_7$~\cite{MNTT05}, the best refinement was obtained using an ordered spin-ice 2-in/2-out structure for moments on a unit tetrahedron (inset of Fig.~\ref{fig_spins}), yielding an on-site moment~$\mu_\mathrm{neu}=1.7(1)\mu_B$ per Pr$^{3+}$ ion~\cite{WMSE15u}. The  ordered spin-ice structure is predicted for long-range ordering of Heisenberg spins on the pyrochlore lattice due to dipole-dipole interactions~\cite{EnGi03u}, although in Pr$_2$Ir$_2$O$_7$\ the Ising nature of the Pr$^{3+}$ moments and the strong dependence of the ordering on stoichiometry suggest RKKY interactions also play an important role.

\begin{figure}[ht]
\includegraphics[width=3.25in]{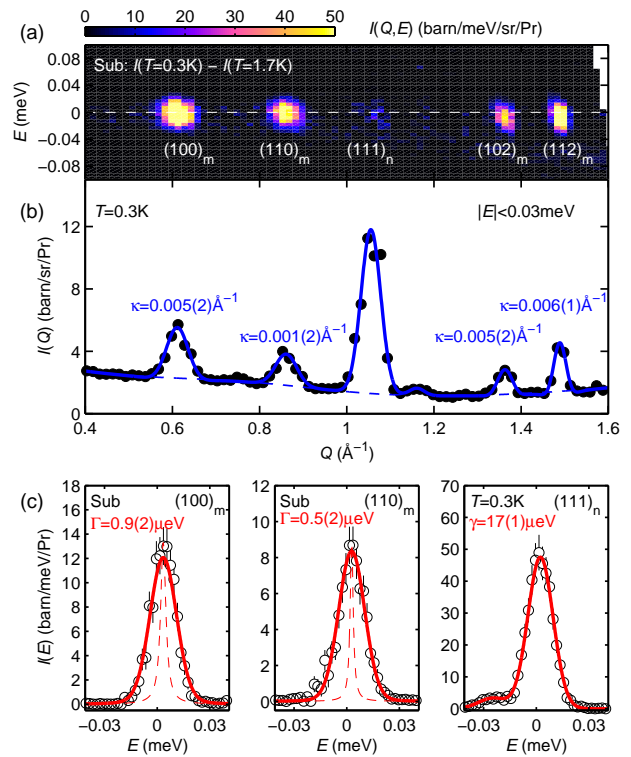}
\caption{(color online) Elastic and quasielastic neutron scattering intensity of Pr$_{2+x}$Ir$_{2-x}$O$_{7-\delta}$ measured at 0.3~K on CNCS\@, $T = 1.7$~K data subtracted. See text for definitions. (a)~Scattering intensity as a function of momentum and energy, $\lambda = 7.26$~\AA. (b)~Momentum dependence of the energy-integrated ($|E|<0.03$~meV) intensity at $T = 0.3$~K\@. Curve: fit to set of Voigt profiles plus a polynomial background. (c)~Energy dependence at three Bragg positions, $\lambda = 9.04$~\AA\@. Solid curves: fits to Voigt profiles. Dashed curves: associated Lorentzian broadening.}
\label{fig_cncs}
\end{figure}

To better understand the spatial and temporal coherence of magnetism below the critical temperature $T_M$, we now turn to high-resolution magnetic neutron scattering. The momentum dependence of the high-temperature-subtracted scattering data [Fig.~\ref{fig_cncs}(a)] reveals four magnetic Bragg peaks, indexed by $(100)$, $(110)$, $(102)$ and $(112)$, that appear sharp in both momentum and energy. 
A fit to the 0.3~K data integrated over $|E| < 0.03$~meV [Fig.~\ref{fig_cncs}(b)] yields a Gaussian momentum resolution of FWHM 0.023(1)~\AA$^{-1}$ at the $(111)$ nuclear Bragg peak. Using a phenomenological expression for the momentum dependence of the momentum resolution, we fit the data to a set of Gaussian-convoluted Lorentzian profiles. This yields the intrinsic half-width-half-maximum (HWHM) widths $\kappa$ for each magnetic Bragg peak in Fig.~\ref{fig_cncs}(b). From this analysis we obtain a lower bound~$\xi_{\rm min} = 1/\kappa_{\rm max} \approx 170$~\AA\ for the spatial correlation length.

The energy dependence of the two lowest-angle magnetic Bragg peaks, measured with $\lambda = 9.04$~\AA, is compared to that of the resolution-limited nuclear Bragg peak $(111)$ in Fig.~\ref{fig_cncs}(c). A fit of the $(100)$ and $(110)$ magnetic Bragg peaks to a quasielastic Lorentzian profile convoluted with a fixed Gaussian energy resolution (FWHM $\gamma=17(1)~\mu$eV) yields intrinsic HWHM widths $\Gamma = 0.9(2)~\mu$eV and 0.5(2)~$\mu$eV, respectively. From this analysis we obtain an upper bound of $\approx1~\mu$eV on any intrinsic broadening, indicating that the observed order is static on a time scale that exceeds $\hbar/\Gamma \approx 0.7$~ns.

Overall our elastic and quasielastic neutron results reveal that our Pr$_{2+x}$Ir$_{2-x}$O$_{7-\delta}$ sample experiences a transition at $T_M = 0.93(1)$~K from a paramagnetic state to long-range spin-ice order characterized by spatial and temporal correlations that span at least $170$~\AA\ and $0.7$~ns, respectively.

\subsection{\label{sec:musr}Muon spin relaxation}

The present $\mu$SR studies of Pr$_2$Ir$_2$O$_7$, like those reported previously~\cite{MOMN09,MNOM10}, were carried out using the dilution refrigerator at the M15 muon beam channel at TRIUMF, Vancouver, Canada. Millimeter-sized crystals were glued to a silver plate with 7031 varnish and mounted on the silver cold finger of the cryostat. $\mu$SR data were taken over the temperature range~25~mK--10~K in weak longitudinal fields~$H_L$ (parallel to the initial $\mu^+$ spin direction, hereafter LF) between 0.4~mT and 2.2~mT that served to prevent muon spin precession in uncompensated stray fields. Data were also collected in LF up to 200~mT at a number of temperatures. All data were analyzed using the Paul Scherrer Institute fitting program~\textsc{musrfit}~\cite{SuWo12}.

\subsubsection{\label{sec:technique}The $\mu$SR technique}

$\mu$SR is a magnetic resonance technique that probes magnetic properties of materials on the atomic scale~\cite{YaDdR11}. Spin-polarized positive muons ($\mu^+$) are implanted in the sample and come to rest at interstitial sites, where magnetic species in the host produce local fields~$\mathbf{B}_\mathrm{loc}(t)$. In Pr$_{2+x}$Ir$_{2-x}$O$_{7-\delta}$ these species are Pr$^{3+}$ $4f$ ions and (predominantly) $^{141}$Pr nuclei:
\begin{equation} \label{eq:Bloc}
\mathbf{B}_\mathrm{loc}(t) = \mathbf{B}_\mathrm{loc}^{4f}(t) + \mathbf{B}^\mathrm{nuc}_\mathrm{loc}(t) \,.
\end{equation}
Both electronic and nuclear contributions vary stochastically with time due to thermal fluctuations. Both contributions are found to be important in $\mathrm{Pr_2Ir_2O_7}$. 

Each muon precesses in the sum of $\mathbf{B}_\mathrm{loc}(t)$ at its site [dynamic relaxation can be thought of as precession in the stochastic time dependence of $\mathbf{B}_\mathrm{loc}(t)$] and any applied field, and eventually decays via the parity-violating weak interaction~$\mu^+ \rightarrow e^+ + \nu_e + \overline{\nu}_\mu$. The direction of the decay positron emission is correlated with the $\mu^+$ spin direction at the time of decay, so that the positron count rate asymmetry (difference between parallel and antiparallel count rates) for a given direction is proportional to the $\mu^+$ spin-polarization component~$P$ in that direction. This property, together with the exponential distribution of $\mu^+$ radioactive decay times (mean lifetime~$= 2.2~\mu$s), allows measurement of the time evolution of the asymmetry~$A(t) = A_0P(t)$ for times up to 10--15~$\mu$s. $A(t)$ is analogous to the free induction signal of NMR~\cite{Abra61,Slic96}. The initial asymmetry~$A_0$ is spectrometer-dependent but is usually $\sim$20\%. Unlike other resonance techniques, e.g., ESR and NMR, $\mu$SR can be carried out in zero and low applied fields.

\subsubsection{\label{sec:rlxproc}Relaxation processes}

In general two classes of relaxation processes contribute to $\mu^+$ spin relaxation (decay of the ensemble $\mu^+$ spin polarization)~\cite{HUIN79,YaDdR11}: 
\begin{itemize}

\item \textit{Quasistatic} relaxation, which is dephasing due to $\mu^+$ precession in a quasistatic~\footnotemark[1] component~$\langle\mathbf{B}_\mathrm{loc} \rangle$ of $\mathbf{B}_\mathrm{loc}(t)$ if the magnitude~$\langle B_\mathrm{loc}\rangle$ is spatially distributed. We denote the quasistatic $\mu^+$ relaxation rate by $\Delta$. The width of the distribution of $\langle B_\mathrm{loc}\rangle$ is $\Delta/\gamma_\mu$, where $\gamma_\mu = 2\pi \times 135.54~\text{MHz~T}^{-1}$ is the muon gyromagnetic ratio.

\item \textit{Dynamic} relaxation, due to the thermally fluctuating component~$\delta\mathbf{B}_\mathrm{loc}(t) = \mathbf{B}_\mathrm{loc}(t) - \langle\mathbf{B}_\mathrm{loc} \rangle$. We denote the dynamic $\mu^+$ relaxation rate by $\lambda$.

\end{itemize}
In NMR these are known as inhomogeneous and homogeneous relaxation, respectively. If a two-component Kubo-Toyabe (K-T)~\cite{KuTo67,HUIN79} structure is observed the two rates can be separated, since then the early-time relaxation is dominated by $\Delta$ and the late time relaxation by $\lambda$ ($\Delta > \lambda$)~\cite{HUIN79,YaDdR11}.

As an example, we consider dipolar fields at $\mu^+$ sites from neighboring nuclear moments in an otherwise nonmagnetic system~\cite{HUIN79}. Nuclear spin dynamics are often quasistatic on the muon time scale, and the field distribution is approximately Gaussian if there are several near-neighbor nuclei~\footnote{The central limit theorem, which is approximately obeyed for a relatively small number of additive random components, yields a Gaussian distribution.}. In zero field (ZF) or LF the relaxation is well described by the static Gaussian K-T function~$G_\mathrm{GKT}(\Delta,\omega_L,t)$~\cite{YaDdR11,KuTo67,HUIN79}, where $\omega_L = \gamma_\mu H_L$. In ZF
\begin{equation} \label{eq:GKT}
G_\mathrm{GKT}(\Delta,t) = {\textstyle\frac{1}{3}} + {\textstyle\frac{2}{3}}[1 - (\Delta t)^2]\exp\left[-{\textstyle\frac{1}{2}}(\Delta t)^2\right] \,;
\end{equation}
the form is more complicated for $\omega_L \ne 0$~\cite{KuTo67,HUIN79}.

In the presence of fluctuating local fields and consequent dynamic relaxation there are several cases, depending on properties of $\delta\mathbf{B}_\mathrm{loc}(t)$. We must consider these cases in detail since, as we show below, it can be difficult to distinguish between them simply on the basis of goodness of fit to the data.

We take $\delta\mathbf{B}_\mathrm{loc}(t)$ to be characterized by its rms amplitude~$\delta B_\mathrm{rms}$ and a correlation time~$\tau_c$ or, equivalently, a fluctuation rate~$\nu = 1/\tau_c$. We define $\delta\omega_\mathrm{rms} = \gamma_\mu\delta B_\mathrm{rms}$. There are then four limiting cases; two concerning the makeup of $\mathbf{B}_\mathrm{loc}(t)$:
\begin{itemize}

\item[1.] $\langle \mathrm{B}_\mathrm{loc}\rangle \ne 0$, i.e., a quasistatic component is present, originating from either nuclear moments or static electronic magnetism, and

\item[2.] $\delta\mathbf{B}_\mathrm{loc}(t) = \mathbf{B}_\mathrm{loc}(t)$, $\langle \mathrm{B}_\mathrm{loc}\rangle = 0$, i.e., $\mathbf{B}_\mathrm{loc}(t)$ reorients fully as a whole;

\end{itemize}
and two concerning the fluctuation frequency:
\begin{itemize}

\item[a.] $\nu \gg \delta\omega_\mathrm{rms}$, the \textit{motionally narrowed} limit, for which $\lambda \approx (\delta\omega_\mathrm{rms})^2\tau_c$~\protect\cite{Abra61,Slic96}, and

\item[b.] $\nu \ll \delta\omega_\mathrm{rms}$, the \textit{adiabatic} limit, for which $\lambda \approx \tau_c$ in ZF.

\end{itemize}
Thus there are four corresponding limiting relaxation types, which we label 1a, 1b, 2a, and 2b in an obvious notation.

In Type~1 (i.e., Type 1a or 1b) relaxation $\delta\mathbf{B}_\mathrm{loc}(t)$ induces transitions between the $\mu^+$ Zeeman levels in the sum of $\langle\mathbf{B}_\mathrm{loc}\rangle$ and any applied field. Assuming a Gaussian distribution of $\langle B_\mathrm{loc}\rangle$, this can be modeled by exponential damping of the static Gaussian K-T relaxation function:
\begin{equation}
P(t) = e^{-\lambda t} G_\mathrm{GKT}(\Delta,\omega_L,t) \,,
\label{eq:EDSKT}
\end{equation}
where the rates~$\Delta$ and $\lambda$ are defined above. We denote fits of Eq.~(\ref{eq:EDSKT}) to the data as exponentially-damped static K-T (EDSKT) fits. 

Type~2 relaxation leads to a number of fitting functions~\cite{KuTo67,HUIN79,YaDdR11}, which depend somewhat on assumptions about the stochastic behavior of $\mathbf{B}_\mathrm{loc}(t)$ but do not differ drastically. We denote such fits as dynamic K-T (DKT) fits since there is no static local field. For Type~2 relaxation $\delta\omega_\mathrm{rms} = \Delta$. A two-component K-T relaxation function is found only in the adiabatic limit (Type~2b relaxation), where the late-time component decays exponentially with rate~$\lambda = \frac{2}{3}\nu$. For Type~2a relaxation the two-component structure is motionally narrowed, and the entire polarization relaxes exponentially with $\lambda = 2\Delta^2/\nu$~\cite{KuTo67,HUIN79}.

\subsubsection{Comparison with experiment}

\begin{figure}[ht]
\includegraphics[clip=,width=3.25in]{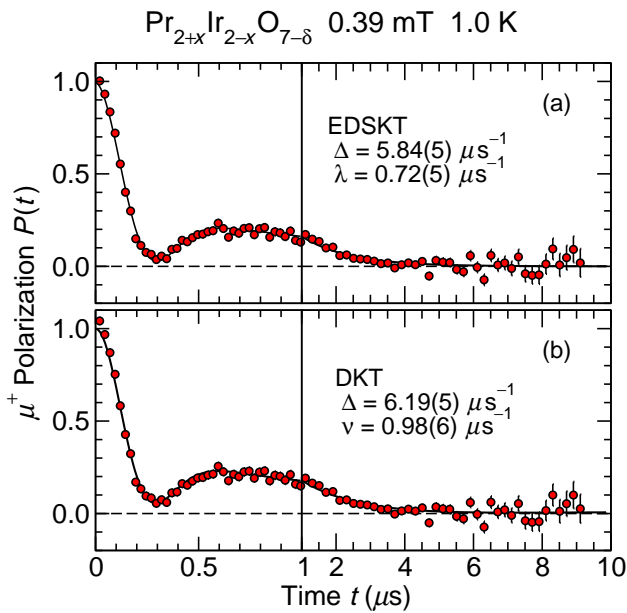}
\caption{\label{fig:6483asy} (color online) Relaxation of $\mu^+$ spin polarization~$P(t)$ in Pr$_{2+x}$Ir$_{2-x}$O$_{7-\delta}$ at $T = 1.0$~K and weak longitudinal field~$\mu_0H_L = 0.39$~mT, obtained from asymmetry data. (a)~EDSKT fit [Eq.~(\protect\ref{eq:EDSKT})], $\chi^2_\mathrm{red} = 1.187$. (b)~DKT fit (Ref.~\protect\cite{HUIN79}), $\chi^2_\mathrm{red} = 1.191$.}
\end{figure}

Figure~\ref{fig:6483asy} shows the time evolution of the normalized $\mu^+$ spin polarization~$P(t)$ for $T = 1.0$~K, $\mu_0H_L = 0.39$~mT, obtained from asymmetry data by subtracting a background signal from muons that miss the sample and stop in the silver cold finger of the cryostat.
The data exhibit the two-component K-T form discussed above, thus ruling out Type~2a relaxation. Figures~\ref{fig:6483asy}(a) and (b) show EDSKT and DKT fits, respectively. The quality of these fits is comparable, so that it is difficult to distinguish between them on that basis, contrary to a previous report~\cite{MNOM10}. 

The quasistatic relaxation rates~$\Delta$ extracted from the fits are in reasonable agreement. The expected relation~$\lambda = \frac{2}{3}\nu$ is roughly obeyed by the fits, but this is due to properties of the fit functions and is not evidence for Type~2b relaxation. There is evidence, discussed below in Sec.~\ref{sec:type?}, that the relaxation is Type~1. Henceforth only EDSKT fits are shown, although parameter values from both EDSKT and DKT fits are compared in Sec.~\ref{sec:tempdep}.

\begin{figure}[ht]
\includegraphics[clip=,width=3.25in]{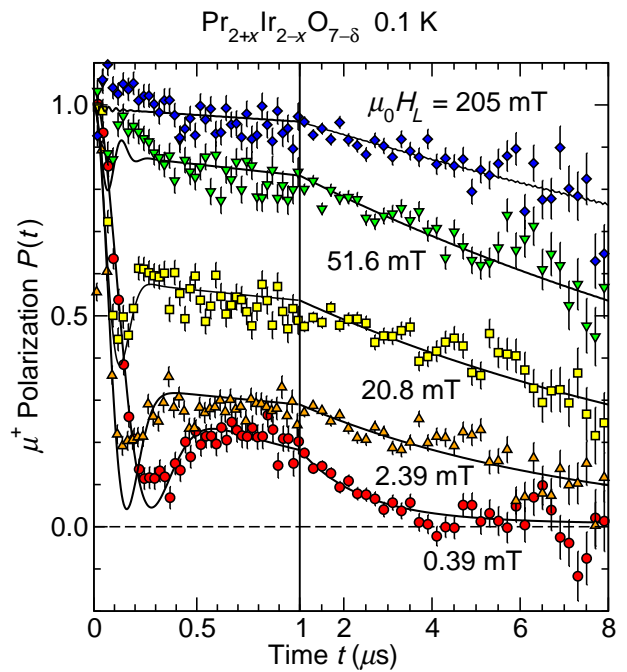}
\caption{\label{fig:6471-7920asy} (color online) 
Relaxation of $\mu^+$ spin polarization~$P(t)$ in Pr$_{2+x}$Ir$_{2-x}$O$_{7-\delta}$ at various longitudinal fields~$H_L$, $T = 0.1$~K\@. Curves: EDSKT fits.}
\end{figure}

It should be noted that a two-component relaxation function might arise from the presence of a spurious phase in the sample. Evidence against this in the present data comes from the dependence of $P(t)$ on longitudinal field~$H_L$, shown in Fig.~\ref{fig:6471-7920asy} for $T = 0.1$~K\@. 
The nearly field-independent rapid relaxation at early times and the increase of the late-time polarization with field are specific signatures of K-T relaxation~\cite{HUIN79}, and would not be expected from a spurious phase. There is some indication of a rapidly-relaxing component, $\sim$10\% of the total polarization, in the high-field data (Fig.~\ref{fig:6471-7920asy}), which probably does arise from one or more spurious phases. 

The late-time data of Fig.~\ref{fig:6471-7920asy} yield the field dependence of the dynamic muon spin relaxation rate. This has also been measured at 25~mK, 0.3~K, and 1~K, and is discussed below in Sec.~\ref{sec:dynamic}.

\section{\label{sec:disc}Results and Discussion}

\subsection{\label{sec:Bonville}Specific heat, elastic neutron scattering} 

As discussed in Sec.~\ref{sec:spht}, the nuclear hyperfine Schottky anomaly in the specific heat is fit by the same value of the ordered Pr$^{3+}$ moment as found from the neutron Bragg scattering intensity [$\mu_\mathrm{neu} = 1.7(1)\mu_B$/Pr ion, Sec.~\ref{sec:neu}], but with a reduced amplitude~$f = 0.65(1)$. This agreement is difficult to understand if $1-f$ is the fraction of Pr$^{3+}$ ions that are not ordered, since then the neutron scattering intensity would be decreased~\footnote{In a diluted magnet with long-range order of moments~$\mu$ the neutron Bragg scattering intensity is is $\propto f\mu^2$. In that case $\mu_\mathrm{neu} = \sqrt{f}\mu_S = 1.3(2)\mu_\mathrm{B}$/Pr ion instead of the observed agreement $\mu_\mathrm{neu} = \mu_S$.}. 

Bertin \textit{et al.}~\cite{BBBH02} showed that a reduced nuclear Schottky specific heat~$C_\mathrm{nuc}$ is expected in a magnetic solid if the ionic moments are fluctuating, i.e., if the system is a cooperative paramagnet. The corresponding fluctuations in the hyperfine field~$B_\mathrm{hf}$ reduce the $^{141}$Pr Zeeman level population differences, and lead to a reduction of $C_\mathrm{nuc}$ when the correlation time~$\tau_c$ of the fluctuations is of the order of or less than the nuclear spin-lattice relaxation time~$T_1$ associated with processes that maintain thermal equilibrium in the nuclear spin system. 

For nuclear spin~1/2 the shape of the Schottky anomaly and the position of the maximum are unchanged to a good approximation (a few \%); only the amplitude is reduced~\cite{BBBH02}:
\begin{equation} \label{eq:Bonville}
C_\mathrm{nuc} \approx f C_\mathrm{nuc}^0, \quad \mathrm{where} \quad f = \frac{1}{1 + 2T_1/\tau_c}
\end{equation}
and $C_\mathrm{nuc}^0$ is the nuclear Schottky specific heat in the absence of fluctuations~\cite{Loun67}. If $T_1 \ll \tau_c$ the fluctuations are unimportant, whereas if $T_1 \gg \tau_c$ the nuclei are always out of equilibrium and do not contribute to the specific heat. Equation~(\ref{eq:Bonville}) remains valid for arbitrary nuclear spin as long as $T_1$ is well defined, i.e., the nuclear spin system is characterized by a spin temperature~\cite{Abra61}~\footnote{For nuclear spin 1/2 relaxation dynamics always lead to a single relaxation time~$T_1$, cf.\ Refs.~\protect\cite{Abra61} and \cite{Slic96}.}. The observed value of $f$ yields $T_1/\tau_c = 0.27(1)$. 

Thus the reduced nuclear Schottky anomaly suggests fluctuating moments rather than static spin freezing in the ordered state of Pr$_{2+x}$Ir$_{2-x}$O$_{7-\delta}$. The $\mu$SR results discussed below in Sec.~\ref{sec:dynamic} are consistent with this picture and place an upper bound of a few nanoseconds on $\tau_c$, slightly longer than the lower bound from elastic neutron scattering. 

\subsection{\label{sec:muSR}Muon spin relaxation}

\subsubsection{\label{sec:quasistatic}Quasistatic relaxation}

\begin{figure}[ht]
\includegraphics[clip=,width= 3.25in]{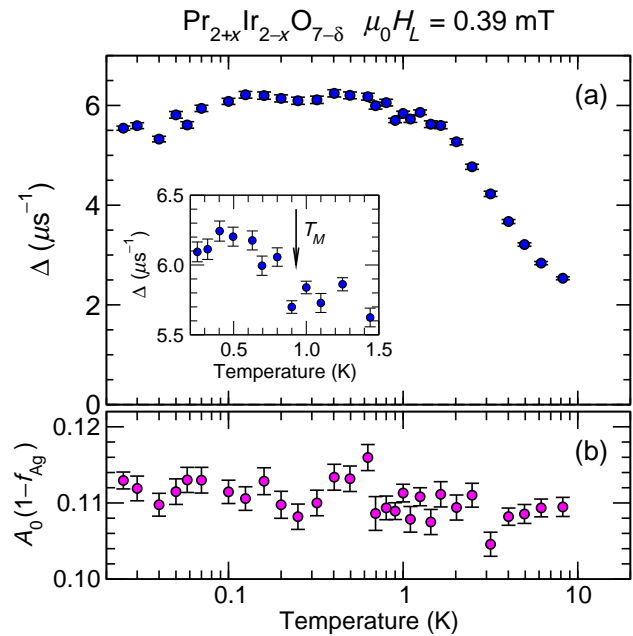}
\caption{\label{fig:DeltaofT} (color online) Temperature dependence of (a)~quasistatic $\mu^+$ relaxation rate~$\Delta$ and (b)~initial ($t = 0$) sample asymmetry~$A_0(1-f_\mathrm{Ag})$ (Sec.~\protect\ref{sec:musr}) from EDSKT fits to weak-LF $\mu$SR data from Pr$_{2+x}$Ir$_{2-x}$O$_{7-\delta}$. Inset to (a): data in the vicinity of the magnetic ordering temperature~$T_M = 0.93$~K from elastic neutron scattering.}
\end{figure}

\paragraph{\label{sec:hfenhance}Hyperfine-enhanced $^{141}$Pr nuclear magnetism.} Figure~\ref{fig:DeltaofT} gives the temperature dependence of the quasistatic $\mu^+$ spin relaxation rate~$\Delta$ for Pr$_{2+x}$Ir$_{2-x}$O$_{7-\delta}$ from EDSKT fits to the data. 
Several features are evident: 
\begin{itemize}

\item $\Delta$ is a fairly smooth function of temperature. A broad maximum in $\Delta(T)$ is observed in the range~0.1--1~K, where $\Delta \approx 6~\mu\text{s}^{-1}$. This corresponds to a rms spread in $\langle B_\mathrm{loc} \rangle$ of roughly 7~mT, more than an order of magnitude greater than expected from the (predominantly $^{141}$Pr) nuclear dipolar contribution to $\mathbf{B}_\mathrm{loc} $. 

\item Above 2~K $\Delta$ decreases markedly with increasing temperature, consistent with the proportionality of $\Delta$ to the local magnetic susceptibility~$\chi_\mathrm{loc}$ found previously~\cite{MNOM10}. 

\item There is a possible small anomaly at the magnetic phase transition temperature~$T_M = 0.93$~K where elastic neutron scattering shows the onset of long-range magnetic order (Fig.~\ref{fig_spins}). 

\end{itemize}
Similar large low-temperature values of $\Delta$ and a smooth temperature dependence (no large anomaly at $T_M$) have been reported previously in other samples of Pr$_2$Ir$_2$O$_7$~\cite{MOMN09,MNOM10} and other Pr-based pyrochlore oxides~\cite{FLML15}. 

As noted in the Introduction, these properties have been attributed~\cite{MOMN09,MNOM10} to hyperfine-enhanced nuclear magnetism (HENM)~\cite{Blea73,*Blea90}, which results from $\mu^+$-induced splitting of the non-Kramers CEF ground states of near-neighbor Pr$^{3+}$ ions. HENM is due to the Van Vleck-like admixture of CEF excited states into a singlet ground state by the nuclear hyperfine interaction, which leads to ``clothing'' of the nuclear moment by $4f$ magnetism. This results in enhancement of the effective nuclear moment by a factor that can be quite large. An enhanced $^{141}$Pr nuclear dipolar field accounts for the strong quasistatic $\mu^+$ spin relaxation. Foronda \textit{et al.}~\cite{FLML15} have observed HENM in a number of Pr-based pyrochlores, and report a density-functional calculation of the $\mu^+$-induced lattice distortion that agrees with the magnitude of the effect. 

Thus the largest contribution to $\langle\mathbf{B}_\mathrm{loc}\rangle$ at $\mu^+$ sites (and the only appreciable contribution above $T_M$) is the enhanced nuclear dipolar field from $^{141}$Pr HENM\@. For the nuclear contribution~$\Delta_\mathrm{nuc}$ to $\Delta$ at low temperatures we take the average~$6.0~\mu\text{s}^{-1}$ of the EDSKT and DKT fit values at $T = 1.0$~K (Fig.~\ref{fig:6483asy}). This is above $T_M$, where only nuclear spin fluctuations are expected to be quasistatic. 

The estimate~$K_\Delta = {\Delta_\mathrm{nuc}}/\Delta_\mathrm{nuc}^0$ of the enhancement factor~$K$ is obtained from the experimental value of ${\Delta_\mathrm{nuc}}$ and calculations of the unenhanced $\mu^+$ relaxation rate~$\Delta_\mathrm{nuc}^0$ due to $^{141}$Pr nuclear dipolar fields at the $\mu^+$ sites~\cite{HUIN79}. In Pr$_{2+x}$Ir$_{2-x}$O$_{7-\delta}$ these sites are not known. We have therefore calculated $\Delta_\mathrm{nuc}^0$ for three assumed $\mu^+$ sites. One is simply a site of high symmetry, and the other two are found from calculations on related pyrochlore compounds~\cite{Duns00,Blun14pc}. We shall see that although results for these sites differ, the differences do not affect qualitative conclusions.

The cubic space group for the $A_2B_2\mathrm{O}_7$ pyrochlore structure is $Fd\overline{3}m$ (No.~227), with the rare earth on the $A$ lattice site. Using Origin Choice~2~\cite{HSW84}, the assumed sites and their Wyckoff positions are 
\begin{itemize}

\item[(a)] $\frac{1}{8} ~\frac{1}{8} ~\frac{1}{8}\ (8a)$ (centers of $B$-atom tetrahedra), 

\item[(b)] $0.16 ~0.16 ~-0.17\ (96g)$ (minimum energy location in Y$_2$Mo$_2$O$_7$ found using Ewald's method~\cite{Duns00}), and 

\item[(c)] $-0.0125 ~0.0471 ~0.2028\ (192i)$ (location in Pr$_2$Sn$_2$O$_7$ from a density functional calculation~\cite{Blun14pc}). 

\end{itemize}

Values of $\Delta_\mathrm{nuc}^0$ for the assumed sites from lattice sums for the second moments~\cite{HUIN79} are shown in Table~\ref{tab:henm}, together with a number of quantities derived from $\mu$SR data. 
\begin{table}[ht]
\caption{\label{tab:henm}
Quantities associated with assumed $\mu^+$ sites~(a)--(c) (Sec.~\protect\ref{sec:quasistatic}) and obtained from data (``obs'' column): $\mu^+$ quasistatic relaxation rates~$\Delta_\mathrm{nuc}^0,\ \Delta_\mathrm{nuc}$ due to $^{141}$Pr dipolar fields; estimated enhancement factors~$K_\Delta$; Van Vleck and local susceptibilities; CEF ground-state energy splittings~$E_\mathrm{CEF}$ from susceptibilities; and Pr$^{3+}$-moment dipolar local field~$B_\mathrm{loc}^{4f}$ in the ordered state.}
\begin{ruledtabular}
\begin{tabular}{lcccc}
 muon site & $8a$ & $96g$ & $192i$ & obs \\
\hline
$\Delta_\mathrm{nuc}^0$, $\Delta_\mathrm{nuc}$ ($\mu\text{s}^{-1}$) & 0.084\footnotemark[1] & 0.293\footnotemark[1] & 0.149\footnotemark[1] & 6.0(2)\footnotemark[2] \\
$K_\Delta = {\Delta_\mathrm{nuc}}/\Delta_\mathrm{nuc}^0$ & 71 & 20 & 40 & $\cdots$ \\
$\chi_\mathrm{VV}$, $\chi_\mathrm{loc}$ (emu/mol Pr) & 0.379\footnotemark[3] & 0.109\footnotemark[3] & 0.214\footnotemark[3] & 0.15(3)\footnotemark[4] \\
$E_\mathrm{CEF}/k_B$ (K)\footnotemark[5] & 10.1 & 35 & 18 & 26(5)\footnotemark[6] \\
$B_\mathrm{loc}^{4f}$ (mT/$\mu_B$) & 88.8\footnotemark[7] & 163.3\footnotemark[7] & 72.6\footnotemark[7] & 1.5(4)\footnotemark[8] \\
\end{tabular}
\footnotetext[1]{Unenhanced $\Delta_\mathrm{nuc}^0$ from second-moment lattice sums.}
\footnotetext[2]{Observed $\Delta_\mathrm{nuc}$ from average of EDSKT and DKT fit values (Fig.~\protect\ref{fig:6483asy}).}
\footnotetext[3]{$\chi_\mathrm{VV}$ from $K_\Delta$ and hyperfine coupling~\protect\cite{Blea73}.}
\footnotetext[4]{$\chi_\mathrm{loc}$ from $\mu^+$ Knight shift~\protect\cite{MNOM10}.}
\footnotetext[5]{$E_\mathrm{CEF}$ from susceptibilities~\protect\cite{VVle32,TAGG97}.}
\footnotetext[6]{Assuming $\chi_\mathrm{loc}$ is a Van Vleck susceptibility.}
\footnotetext[7]{$B_\mathrm{loc}^{4f}$ from lattice sums for full ordered lattice (no $\mu^+$-induced suppression).}
\footnotetext[8]{Upper bound on $B_\mathrm{loc}^{4f}$ from data [inset to Fig.~\protect\ref{fig:DeltaofT}(a)] assuming $1.7\mu_B$/Pr ion.}
\end{ruledtabular}
\end{table}
The right-hand column~``obs'' gives observed quantities to be compared with the calculations. The bottom row in Table~\ref{tab:henm} concerns the $\mu^+$ local dipolar field~$B_\mathrm{loc}^{4f}$ from Pr$^{3+}$ moments in the ordered state, and is discussed below in Sec.~\ref{sec:magord}.

We note the following: 
\begin{itemize}

\item The values of $\Delta_\mathrm{nuc}^0$ vary considerably from site to site, as expected since they are dominated by dipolar fields from near-neighbor $^{141}$Pr ions with different distances and spatial configurations. This variation is reflected in the related quantities discussed below. 

\item The values of $K_\Delta$ are all considerably larger than 1, and vary between sites because of the variation of $\Delta_\mathrm{nuc}^0$.

\item The HENM enhancement factor~$K$ is given by $K = a_\mathrm{hf} \chi_\mathrm{VV}$, where $a_\mathrm{hf} = 187.7$~mol/emu is the $^{141}$Pr atomic hyperfine coupling constant and $\chi_\mathrm{VV}$ is the molar Van Vleck susceptibility~\cite{MNOM10,Blea73}. Values of $\chi_\mathrm{VV}$ obtained from this relation and $K_\Delta$ are comparable to the $\mu^+$-perturbed local susceptibility~$\chi_\mathrm{loc}$ from the $\mu^+$ Knight shift. This shows consistency of the $\mu$SR measurements.

\item The CEF splitting energies~$E_\mathrm{CEF}$ calculated from $\chi_\mathrm{VV}$ and $\chi_\mathrm{loc}$~\cite{VVle32} are comparable to measured $\mu^+$-induced CEF energy shifts in a number of Pr-based compounds~\cite{TAGG97}. 

\end{itemize}

\paragraph{\label{sec:magord}Static magnetic order?} Magnetic order usually results in an increase in $\Delta$ [or, if $\langle B_\mathrm{loc} \rangle$ is homogeneous, oscillations in $P(t)$] below the transition temperature. Assuming that the small increase in $\Delta$ below $T_M$ [inset of Fig.~\ref{fig:DeltaofT}(a)] is due to magnetic order, we consider the magnitude of that order. 

The dominant contribution to $\Delta$ below $T_M$ is $\Delta_\mathrm{nuc}$, which is due to random nuclear dipolar fields. Thus $\Delta_\mathrm{nuc}$ and any additional static relaxation rate or precession frequency~$\Omega_\mathrm{mag}$ add in quadrature, so that $\Omega_\mathrm{mag} = \sqrt{\Delta^2 - \Delta_\mathrm{nuc}^2}$. From the small anomaly in $\Delta(T)$ [inset to Fig.~\ref{fig:DeltaofT}(a)] $\Omega_\mathrm{mag} = 2.2(6)~\mu\text{s}^{-1}$. The anomaly at $T_M$ is not very significant statistically, however, and this value should be considered as an upper bound on $\Omega_\mathrm{mag}$ rather than an established result. In a previously-studied sample~\cite{MNOM10} with evidence for a transition at 0.8~K from bulk properties, the upper bound is even smaller. Then the Pr$^{3+}$-moment contribution~$\langle B_\mathrm{loc}^{4f}\rangle$ is $\Omega_\mathrm{mag}/\gamma_\mu = 2.6(7)$~mT, or 1.5(4)~mT/$\mu_\mathrm{B}$ (Table~\ref{tab:henm}), assuming an ordered moment~$\mu_\mathrm{Pr} = 1.7(1)\mu_\mathrm{B}$/Pr ion from the nuclear Schottky specific heat and neutron Bragg scattering results (Secs.~\ref{sec:spht} and \ref{sec:neu})\@.

We have carried out dipolar-field lattice sums of $B_\mathrm{loc}^{4f}$ for spin-ice (2-in/2-out) ordered moments in Pr$_2$Ir$_2$O$_7$, assuming the three $\mu^+$ sites discussed above. For the full lattice (no $\mu^+$-induced moment suppression) these calculations yield $B_\mathrm{loc}^{4f} \approx $70--160~mT/$\mu_\mathrm{B}$ (Table~\ref{tab:henm}), nearly two orders of magnitude larger than $\Omega_\mathrm{mag}/\gamma_\mu $. Thus if all Pr moments were ordered uniformly, including nearest neighbors to the $\mu^+$ site, the $\mu$SR data indicate an ordered moment~$\lesssim 0.02\mu_\mathrm{B}$/Pr ion. This small value disagrees strongly with the nuclear Schottky anomaly and elastic neutron scattering results.

As noted in the Introduction, a possible cause of the reduced Pr-moment contribution to $B_\mathrm{loc}^{4f}$ is the $\mu^+$-induced suppression of near-neighbor Pr moments that gives rise to the observed HENM\@. In Pr$_2$Ir$_2$O$_7$ the Pr$^{3+}$ ground-state doublet consists of almost pure $|J_z{=}{\pm}4\rangle$ states~\cite{MNTT05,Mach07}. It is straightforward to show that any matrix element between these states admixes them equally. There is no such matrix element of the CEF in the trigonal point symmetry at the Pr$^{3+}$ lattice site; this is the origin of the non-Kramers doublet ground state. The $\mu^+$-induced lattice distortion breaks this symmetry, splits the non-Kramers doublet as required for HENM, and suppresses the ground-state moment~\cite{MOMN09,MNOM10,FLML15}. 

\begin{figure}[ht]
\includegraphics[clip=,width=3.25in]{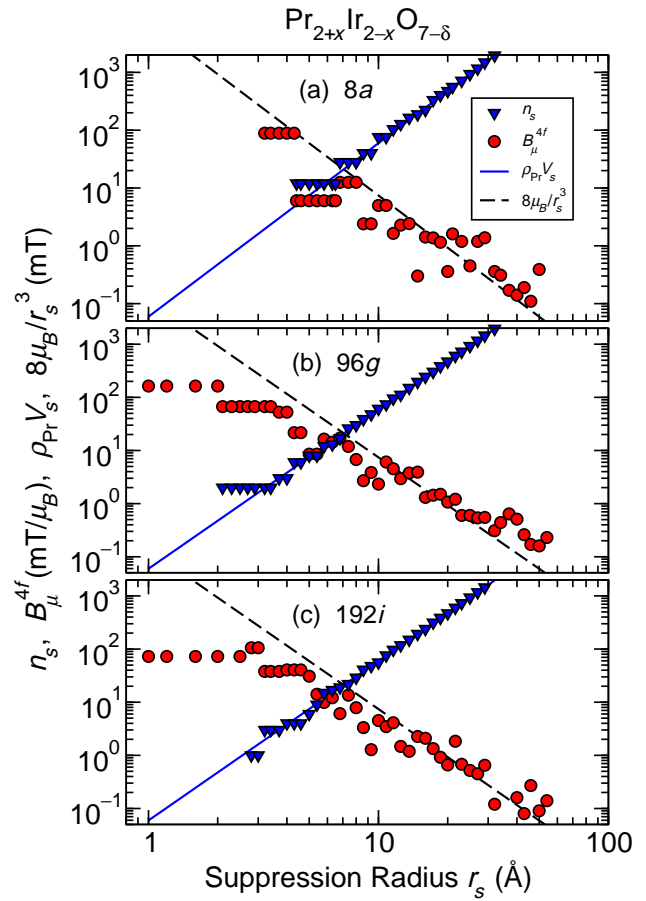}
\caption{\label{fig:dipsum}(color online) Lattice sums for the $\mu^+$-induced suppressed-moment hypothesis in Pr$_{2+x}$Ir$_{2-x}$O$_{7-\delta}$ as functions of suppression radius~$r_s$ at three assumed $\mu^+$ sites (see text). Triangles:~number~$n_s$ of suppressed-moment sites. Circles: Pr$^{3+}$ $4f$ dipolar field~$B_\mathrm{loc}^{4f}$ at $\mu^+$ site assuming the ordered spin-ice structure. Solid lines:~$\rho_\mathrm{Pr} V_s$, where $\rho_\mathrm{Pr}$ is the Pr-site number density and $V_s = (4\pi/3)r_s^3$ is the suppression volume. Dashed lines: $8\mu_\mathrm{B}/r_s^{-3}$ (see text).} 
\end{figure}

Dipolar-field lattice sums for $B_\mathrm{loc}^{4f}$ have also been carried out with Pr$^{3+}$ $4f$ moments suppressed to zero within a radius~$r_s$ and corresponding volume~$V_s = (4\pi/3)r_s^3$ around each of the three assumed $\mu^+$ sites. The ordered spin-ice structure is assumed outside this volume. The results are shown in Fig.~\ref{fig:dipsum} 
as functions of $r_s$, where the quantities plotted are: the number~$n_s$ of suppressed Pr$^{3+}$ moments, the Pr$^{3+}$ $4f$ dipolar field~$B_\mathrm{loc}^{4f}$ at the $\mu^+$ site, $\rho_\mathrm{Pr} V_s$, where $\rho_\mathrm{Pr} = 1.32 \times 10^{22}~\text{cm}^{-3}$ is the Pr-ion number density ($n_s \to \rho_\mathrm{Pr}V_s$ for large $r_s$), and the quantity $8\mu_\mathrm{B}/r_s^3$ for comparison. The leftmost values of $B_\mathrm{loc}^{4f}$ are for $n_s = 0$ ($r_s$ smaller than nearest-neighbor distance) in all three panels. As $r_s$ increases $B_\mathrm{loc}^{4f}$ changes in jumps, with an overall decrease that is given roughly by $8\mu_\mathrm{B}/r_s^3$ for large $r_s$. This indicates that $\sim$8 effective moments remain on the surface of the suppressed-moment sphere. It is not surprising that the qualitative behavior does not depend strongly on $\mu^+$ site for $r_s \gtrsim 7$~\AA. 

The suppression region required to reduce the dipolar field from the remaining moments to $\lesssim 1.5~\mathrm{mT}/\mu_\mathrm{B}$ is large: $r_s \approx 14$~\AA, $n_s \approx 160$ Pr ions. Although the range of CEF perturbation around a point-charge impurity has to our knowledge not been determined, previous results~\cite{TAGG97,FLML15} suggest significant effects for only a few Pr near neighbors. If this is the case, a large value of $r_s$ would have to be due to a long ``healing length'' for the magnetic order, rather than to the lattice distortion itself, i.e., the majority of suppressed Pr$^{3+}$ $4f$ moments would have to be suppressed by suppressed Pr$^{3+}$ $4f$ neighbors. To our knowledge such a phenomenon has not been reported previously. Indeed, moment suppression would be in contrast to the impurity-induced magnetism often found in paramagnetic systems such as frustrated magnets that are close to magnetic instabilities~\cite{PRYP98,TSH99,TaHu02,XABB00}. We note in passing that disorder-induced magnetism would be consistent with the observation of magnetic order only in stuffed samples.

If the $\mu^+$ relaxation rate is due to disordered static magnetism and is fast, a fraction of the $\mu^+$ asymmetry can be lost in the spectrometer ``dead'' time. The initial sample asymmetry~$A_0(1-f_\mathrm{Ag})$ from EDSKT fits (cf.\ Sec.~\ref{sec:musr}), shown in Fig.~\ref{fig:DeltaofT}(b), exhibits no such loss below $T_M$, thereby ruling out very rapid relaxation.

\subsubsection{\label{sec:dynamic} Dynamic relaxation} 

\begin{figure}[ht]
\includegraphics[clip=,width=3.25in]{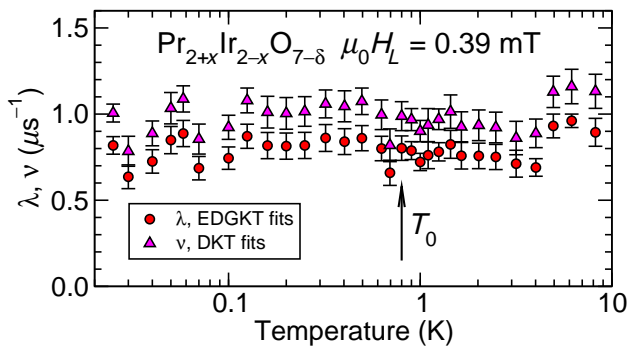}
\caption{\label{fig:lambdaofT} (color online) Temperature dependence of $\mu^+$ dynamic relaxation parameters in Pr$_{2+x}$Ir$_{2-x}$O$_{7-\delta}$ from fits to weak-LF $\mu$SR data. Circles: $\lambda(T)$ from EDSKT fits. Triangles: $\nu(T)$ from DKT fits.}
\end{figure}

\paragraph{\label{sec:tempdep}Temperature dependence.} Figure~\ref{fig:lambdaofT} shows the temperature dependence of $\mu^+$ dynamic relaxation parameters~$\lambda$ from EDSKT fits and $\nu$ from DKT fits. 
The near temperature independence of $\lambda$ is often observed in pyrochlores and other geometrically frustrated systems~\cite{CaKe11}. It is sometimes given as evidence for persistent spin dynamics at low temperatures; this is unexpected, since dynamic relaxation in magnetic resonance experiments is normally driven by thermal fluctuations.

The Pr-Pr nearest-neighbor distance in Pr$_2$Ir$_2$O$_7$ is $\sim$3.8~\AA\@. Assuming a suppression radius~$r_s$ slightly longer than this (5--7~\AA), Fig.~\ref{fig:dipsum} shows that 7--20 Pr moments around the muon sites are suppressed. The remaining bulk moments, assumed to be 1.7$\mu_\mathrm{B}$ in the ordered spin-ice configuration, give dipolar fields~$B_\mathrm{loc}^{4f} = 10$--20~mT at the muon site. The spread reflects the dependence of $B_\mathrm{loc}^{4f}$ on the site and the assumed value of $r_s$, and is a conservative estimate of the uncertainty. This yields $\Omega_{4f} = \gamma_\mu B_\mathrm{loc}^{4f} \approx 10\text{--}20~\mu\text{s}^{-1}$, of the order of $\Delta$ and an order of magnitude greater than the observed~$\lambda \simeq 1~\mu\text{s}^{-1}$. This is the magnitude of the fluctuating $\mu^+$ local field in the fluctuating-moment picture.

\paragraph{\label{sec:type?}Which type of dynamic relaxation?}We wish to determine whether dynamic relaxation in Pr$_{2+x}$Ir$_{2-x}$O$_{7-\delta}$ is Type~1 or Type~2b (cf.\ Sec.~\ref{sec:rlxproc}), i.e., whether EDSKT fits or adiabatic-limit DKT fits are most appropriate. We have seen (Sec.~\ref{sec:musr}, Fig.~\ref{fig:6483asy}) that this cannot be done from the goodness of fit. 

We first consider Type~2b relaxation. Enhanced $^{141}$Pr nuclear moments dominate $\mathbf{B}_\mathrm{loc} $, so that in this case (which by definition involves complete reorientations of $\mathbf{B}_\mathrm{loc} $) $\mu^+$ relaxation must be due to Pr nuclear spin dynamics. The nuclear spin-spin relaxation rate~$1/T_2$ is normally temperature-independent~\footnote{Ref.~\protect\cite{Abra61}, chapter 3}, so that the enhanced $^{141}$Pr rate~$1/T_2^\mathrm{nuc}$ is a natural candidate for the observed DKT-fit dynamic fluctuation rate~$\nu$ (Fig.~\ref{fig:lambdaofT}). 

The unenhanced $^{141}$Pr rate~$(1/T_2^\mathrm{nuc})_0$ is found to be $8.9 \times 10^{-3}~\mu\text{s}^{-1}$ from a lattice sum for the dipolar second moment~\cite{Abra61,Slic96}. The enhanced $^{141}$Pr rate is given approximately by $1/T_2^\mathrm{nuc} \approx K^{2}/(T_2^\mathrm{nuc})_0$~\cite{Blea73}, which yields an enhancement~$K_{T_2} \approx [\nu(T_2^\mathrm{nuc})_0]^{1/2} \approx 10$ from the DKT fit assuming $1/T_2^\mathrm{nuc} = \nu$. Table~\ref{tab:T2} gives the values of $1/T_2^\mathrm{nuc}$ enhanced by $K_\Delta$ for the assumed $\mu^+$ sites discussed in Sec.~\ref{sec:quasistatic}, together with the observed value of $\nu$ from the DKT fit of Fig.~\ref{fig:6483asy}(b). 
\begin{table}[ht]
\caption{\label{tab:T2} Calculated enhanced $^{141}$Pr spin-spin relaxation rate~$1/T_2^\mathrm{nuc} = K_\Delta^2/(T_2^\mathrm{nuc})_0$ associated with assumed $\mu^+$ sites~(a)--(c) (Sec.~\protect\ref{sec:quasistatic}). ``Observed'' (obs) column: DKT-fit fluctuation rate~$\nu$, $T = 1.0$~K [Fig.~\protect\ref{fig:6483asy}(b)].}
\begin{ruledtabular}
\begin{tabular}{lcccc}
 muon site & $8a$ & $96g$ & $192i$ & obs \\
\hline
$1/T_2^\mathrm{nuc},\ \nu\ (\mu\text{s}^{-1})$ & 44 & 3.7 & 14.3 & 0.98 \\
\end{tabular}
\end{ruledtabular}
\end{table}
It can be seen that the $1/T_2^\mathrm{nuc}$ enhanced by $K_\Delta$ for the assumed $\mu^+$ sites are faster, and except for the $96g$ site considerably faster, than $\nu$. The enhanced $1/T_2^\mathrm{nuc}$ is of the order of or faster than $\Delta$. As noted in Sec.~\ref{sec:rlxproc}, this would lead to a single exponential Type~2a relaxation function~\cite{KuTo67,HUIN79} rather than the observed two-component function (Fig.~\ref{fig:6483asy}). The origin of this apparent overestimate of $1/T_2^\mathrm{nuc}$ is probably the strong spatial inhomogeneity of the $\mu^+$-induced HENM, which limits the ability of neighboring $^{141}$Pr dipolar fields to undergo mutual spin flips~\cite{Abra61,Slic96}. 

Assuming Type-2 relaxation, it seems unlikely that the fluctuation rate~$\nu$ is dominated by an enhanced $1/T_2^\mathrm{nuc}$. $\Delta(T)$, and therefore the enhancement factor~$K_\Delta(T)$, decrease by a factor of $\sim$3 with increasing temperature in the range~1--10~K (Fig.~\ref{fig:DeltaofT}). Thus the enhanced $1/T_2^\mathrm{nuc}(T)$, which is expected to be proportional to $K_\Delta^2(T)$~\cite{Blea73}, should decrease by a factor of $\sim$10 over this temperature range. In contrast, the observed $\nu(T)$ (Fig.~\ref{fig:lambdaofT}) is constant or perhaps increases slightly near 10~K. 

The only other mechanism that could lead to Type~2a $\mu^+$ dynamic relaxation is spin-lattice relaxation of near-neighbor $^{141}$Pr nuclei. But this would be expected to exhibit considerable temperature dependence, especially in the neighborhood of the magnetic phase transition. 

Thus the evidence suggests that the relaxation is Type~1. This in turn suggests a dynamic component of $\mathbf{B}_\mathrm{loc} $ due to Pr$^{3+}\ 4f$ moment fluctuations, since $^{141}$Pr nuclear spins by themselves are unlikely to exhibit fast and slow fluctuations simultaneously.

We note that the onset of quasistatic fluctuations of $\mathbf{B}_\mathrm{loc}^{4f}$ ($\tau_c \gg 1/\Omega_{4f}$) at $T_M$ is unlikely because it would yield an additional quasistatic relaxation rate $\approx \Omega_{4f}$; this would be easily visible. Thus the fluctuations are motionally narrowed in the fluctuating-moment picture; the relaxation is Type~1a (Sec.~\ref{sec:rlxproc}). Since $\lambda$ is an upper bound on the rate due to bulk Pr-moment fluctuations (i.e., the value assuming no $^{141}$Pr nuclear contribution to $\lambda$, Sec.~\ref{sec:hfenhance}), we have as a rough upper bound 
\begin{equation}
\tau_c \lesssim \lambda/2\Omega_{4f}^2 \approx 1.2\text{--}5~\text{ns} \,.
\end{equation}
The neutron elastic scattering lower bound on $\tau_c$ of $\sim$1~ns (Sec.~\ref{sec:neu}) is consistent with this result. Together the two bounds yield $\tau_c$ of the order of a few nanoseconds.

\begin{figure}[ht]
\includegraphics[clip=,width=3.25in]{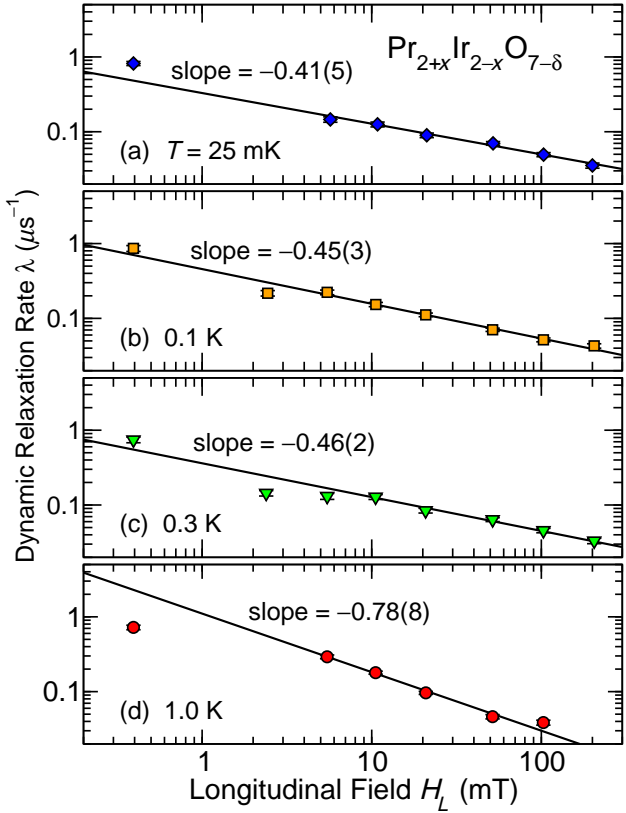}
\caption{\label{fig:lambdaofH} (color online) Dependence of $\mu^+$ dynamic relaxation rate~$\lambda$ on longitudinal magnetic field~$H_L$ from EDSKT fits of Eq.~(\protect\ref{eq:EDSKT}) to weak-LF $\mu$SR data from Pr$_{2+x}$Ir$_{2-x}$O$_{7-\delta}$ at representative temperatures. (a)~$T = 25$~mK\@. (b)~$T = 0.1$~K\@. (c)~$T = 0.3$~K\@. (d)~$T = 1.0$~K\@. Lines: fits of power laws~$\lambda(H_L) \propto H_L^a$ to the data for $\mu_0H_L > 10$~mT.}
\end{figure}

\paragraph{Field dependence.} Figure~\ref{fig:lambdaofH} gives the dependence of $\lambda$ on $H_L$ from EDSKT fits to data at four representative temperatures. 
The straight lines are power-law fits~$\lambda \propto H_L^{-a}$ for $\mu_0H_L > 10$~mT\@. For weaker fields $H_L \lesssim \Delta/\gamma_\mu $ the applied field is a small perturbation on $\langle\mathbf{B}_\mathrm{loc} \rangle$, and little field dependence is expected. It can be seen that the fits are good above 10~mT\@. At 0.1~K and 0.3~K the data level off down to $\sim$2.5~mT, as expected, but for $\mu_0H_L \approx 0.4$~mT the rates are high by comparison. At 0.3~K and below the high-field data are not very temperature dependent, and the power-law exponent~$a$ is roughly constant at $\sim$0.4. In contrast, at $T = 1.0$~K $a$ has nearly doubled, and in 0.4~mT $\lambda$ is $\sim$3 times smaller than the extrapolated power-law fit.

 In the motionally narrowed limit, and in a longitudinal field~$H_L$, the Kubo-Toyabe relaxation rate is of the Redfield form~\cite{HUIN79}
\begin{equation} \label{eq:Redfield}
\lambda = 5\,\Omega_{4f}^2\, \frac{\tau_c}{1 + (\gamma_\mu H_L\tau_c)^2} \,,
\end{equation}
so that for $\gamma_\mu H_L\tau_c \gg 1$ $\lambda \propto (H_L)^{-a}$ with $a = 2$. This value is much larger than the experimental results of Fig.~\ref{fig:lambdaofH}, although the increase of $a$ at 1~K [Fig.~\ref{fig:lambdaofH}(d)] suggests that it might approach 2 at higher temperatures.

A relaxation rate that exhibits a power-law dependence on field is a particular case of time-field scaling~\cite{KMCL96}
\begin{equation} \label{eq:timefield}
P(t,H) = P(t/H^a)
\end{equation}
of the relaxation function~$P(t,H)$, since Eq.~(\ref{eq:timefield}) is obeyed if $P(t,H) = P[(\lambda(H)t]$, $\lambda(H) \propto H^{-a}$. This is the case in Pr$_{2+x}$Ir$_{2-x}$O$_{7-\delta}$. Time-field scaling of a relaxation function has been interpreted as evidence for a zero-frequency divergence of the noise power spectrum associated with ``glassy'' spin dynamics~\cite{KMCL96,KBMC00,MBHN01,Kere04} or proximity to a quantum critical point, with or without structural disorder~\cite{MHBI04,MRAS06}. These associations are valid only if the applied field does not modify the spin dynamics, but merely sweeps the $\mu^+$ precession frequency through the noise power spectrum.

\section{\label{sec:concl}Conclusions}

Muon-induced Pr$^{3+}$ CEF ground-state splitting and HENM dominate the $\mu$SR data in Pr$_{2+x}$Ir$_{2-x}$O$_{7-\delta}$ and complicate their interpretation. Nevertheless, as discussed above in Sec.~\ref{sec:muSR}, the absence of a signature of magnetic order in the $\mu$SR data can be understood as due to one of two novel phenomena (or perhaps to both in some measure): long-range $\mu^+$-induced moment suppression, or nanosecond-scale Pr-moment fluctuations in the magnetically long-range ordered state below $T_M$. There are, however, difficulties with both of these scenarios.

In the moment-suppression picture the suppression volume~$V_s$ must be very large compared to estimates of the region where the $\mu^+$ induces significant CEF level splitting. Then the majority of Pr moments within $V_s$ are suppressed simply by proximity to suppressed-moment neighbors. To our knowledge there is no other evidence, theoretical or experimental, for such an effect. On the contrary, impurities in a number of cooperative paramagnets tend to produce local magnetism rather than suppressing it~\cite{PRYP98,TSH99,TaHu02,XABB00}. 

As noted above (Secs.~\ref{sec:intro} and \ref{sec:Bonville}), Bertin \textit{et al.}~\cite{BBBH02} showed that $4f$ moment fluctuations in a rare-earth-based compound can reduce the amplitude of the nuclear hyperfine Schottky anomaly in the specific heat but leave the temperature dependence unchanged to a good approximation. The magnitude of the fluctuating moment can be determined if the hyperfine interaction is known. In this sample the nuclear Schottky anomaly yields a reduction factor~$f = 0.65(1)$ (Sec.~\ref{sec:spht}), and specific heat and elastic neutron scattering data both yield the same value [1.7(1)$\mu_\mathrm{B}$/Pr ion] for the ordered Pr moment (Secs.~\ref{sec:spht} and \ref{sec:neu}). This would not be the case if the reduced Schottky amplitude were simply due to disorder-induced ground-state splitting and suppression of a fraction $1 - f$ of Pr$^{3+}$ moments, since then the neutron-diffraction value, which is calculated assuming $f = 1$, would not agree with the specific-heat value (cf.\ Secs.~\ref{sec:spht} and \ref{sec:Bonville}).

This agreement does not involve $\mu^+$-induced effects, and thus is independent evidence for the slow-fluctuation scenario. The latter requires spatially correlated fluctuations of regions $\sim$250~\AA\ in size (${\sim}10^6$ Pr moments), however, since this is the minimum correlation length consistent with the elastic neutron scattering results (Sec.~\ref{sec:neu}). Fluctuations of such large correlated regions seem counterintuitive.

Similar behavior has, however, been reported previously in the rare-earth pyrochlore stannates~Gd$_2$Sn$_2$O$_7$ and Tb$_2$Sn$_2$O$_7$~\cite{BBBH02,MAR-CB05,BMS08,Bonv10}. The case closest to that of Pr$_{2+x}$Ir$_{2-x}$O$_{7-\delta}$ seems to be Tb$_2$Sn$_2$O$_7$, where magnetic Bragg peaks are observed~\cite{MMBA08} with a Tb$^{3+}$ moment value markedly larger than that found from the nuclear hyperfine Schottky anomaly in the specific heat~\footnote{Only the $1/T^2$ high-temperature tail of the Schottky anomaly was observed; the discrepancy in ``moment values'' was ascribed to a reduced Schottky amplitude [$f$ in Eq.~(\protect\ref{eq:Bonville})].}. In Tb$_2$Sn$_2$O$_7$, as in Pr$_{2+x}$Ir$_{2-x}$O$_{7-\delta}$, there is no onset of either oscillations or quasistatic $\mu^+$ relaxation at low temperatures in $\mu$SR data~\cite{DdRYKC06}, although there are indirect signs of nearby long-range order~\cite{BMOB06}. It is noteworthy that HENM is not observed in Tb$_2$Sn$_2$O$_7$, even though Tb$^{3+}$ is a non-Kramers ion. A nuclear Schottky anomaly has not been observed in Gd$_2$Sn$_2$O$_7$ due to the smaller Gd$^{3+}$ hyperfine interaction, but $^{155}$Gd M\"ossbauer data indicate a lack of thermal equilibrium associated with a fluctuating hyperfine field~\cite{BBBH02,BMS08,Bonv10}.

The $\mu^+$-induced suppression of neighboring Pr$^{3+}$ moments also complicates the interpretation of the observed muon dynamic relaxation rate~$\lambda$. Nevertheless, it seems unlikely that $\lambda$ is entirely due to $^{141}$Pr spin-spin interactions. These are hard to estimate, particularly in the presence of HENM, but the temperature dependence of the enhancement is not reflected in the fluctuation rate~$\nu$ from DKT fits. It is also difficult to see how the observed power-law field dependence (Fig.~\ref{fig:lambdaofH}) would come about if the $\mu^+$ relaxation were due mainly to $^{141}$Pr nuclear spin fluctuations. On the other hand, one would expect a strong feature in $\lambda$ at $T_M$ from Pr$^{3+}$ moment fluctuations, due to the onset of long-range correlations observed in elastic neutron scattering (Fig.~\ref{fig_spins}). 

If the relaxation is nevertheless dominated by Pr$^{3+}$ $4f$ spin fluctuations, it seems most likely that the motionally-narrowed limit is applicable and the $4f$ correlation time $\tau_c$ is a few nanoseconds. This time scale is slightly beyond the resolution of the neutron scattering experiments to date, but is readily accessed using the neutron spin echo (NSE) technique. An experimental value of $\tau_c$ in the nanosecond range from NSE experiments would be strong evidence in favor of the fluctuating-moment scenario, whereas a longer correlation time (inconsistent with motionally-narrowed $\mu^+$ relaxation) would suggest $\gtrsim 200$ suppressed moments. 

At present it is hard to understand the experimental results in Pr$_{2+x}$Ir$_{2-x}$O$_{7-\delta}$ on either the fluctuating-moment picture or the suppressed-moment picture, although the available evidence leans toward the former. It should be noted, however, that either slow fluctuations of an ordered structure or long-range $\mu^+$-induced moment suppression strongly suggests that Pr$_{2+x}$Ir$_{2-x}$O$_{7-\delta}$ is very close to a magnetic-nonmagnetic critical point. This is perhaps the main conclusion that can be drawn from our results. 

\bigskip \begin{acknowledgments}
C.L.B. and S.N. thank the Aspen Center for Physics, where progress was made on this project, for their hospitality during the summer of 2014. D.E.M. wishes to thank the Institute for Solid State Physics, Tokyo University, for their hospitality during his stays there. We are grateful for technical assistance from the TRIUMF Centre for Molecular and Materials Science, where the $\mu$SR experiments were carried out. We thank E.~J. Ansaldo, J.~M. Mackie, K. Onuma, and S. Zhao for assistance with the experiments, and R.~F. Kiefl and G.~M. Luke for useful discussions. We are grateful to S. Koohpayeh for performing powder X-Ray diffraction at IQM\@. This work was partially supported by U.S. NSF Grant nos.~0422671 and 0801407 (Riverside) and 1105380 (Los Angeles), by a Grant-in-Aid (No.~21684019) from the Japanese Society for the Promotion of Science (JSPS), by Grants-in-Aid for Scientific Research on Priority Areas (Nos.~17071003 and 19052003) from the Ministry of Education, Culture, Sports, Science and Technology (MEXT), Japan, and by the National Natural Science Foundation of China (No.~11474060) and STCSM of China (No.~15XD1500200). The work at IQM was supported by the U.S. Department of Energy, Office of Basic Energy Sciences, Division of Material Sciences and Engineering under grant DE-FG02-08ER46544. The research at ORNL's Spallation Neutron Source was sponsored by the Scientific User Facilities Division, Office of Basic Energy Sciences, U.S. Department of Energy.
\end{acknowledgments}



%

\end{document}